\documentclass[journal]{IEEEtran}

\usepackage{multirow}
\usepackage[table,xcdraw]{xcolor}
\usepackage{graphicx}
\usepackage{amsmath}
\usepackage{amsfonts}
\usepackage{algorithm}
\usepackage{amsthm,amssymb}
\usepackage{mathrsfs}
\usepackage{algorithmic}
\usepackage{makecell}%解决公式在表格中位置太小的问题
\usepackage{easyReview}
\usepackage{booktabs}%解决表格横向粗细问题
\setcellgapes{3pt}
\usepackage{verbatim}
\setcellgapes{3pt}

\usepackage[justification=centering]{caption}
\usepackage{color}
\usepackage[subfigure]{tocloft}
\usepackage{subfigure}
\usepackage{tabularx}
\usepackage{amsmath, bm}
\usepackage{hyperref}
\hypersetup{hidelinks,
	colorlinks=true,
	allcolors=blue,
	pdfstartview=Fit,
	breaklinks=true}
\usepackage{cite}

\ifCLASSINFOpdf
\else
\fi
\begin{document}

\title{Large Generative Model-assisted Talking-face Semantic Communication System}

\author{Feibo Jiang, \textit{Senior Member, IEEE}, Siwei Tu, Li Dong, 
%Kezhi Wang, \textit{Senior Member, IEEE}, Kun Yang, \textit{Fellow, IEEE},
Cunhua Pan, \textit{Senior Member, IEEE}, Jiangzhou Wang, \textit{Fellow, IEEE}, Xiaohu You, \textit{Fellow, IEEE}
	\thanks{
		Feibo Jiang (jiangfb@hunnu.edu.cn) is with Hunan Provincial Key Laboratory of Intelligent Computing and Language Information Processing, Hunan Normal University, Changsha, China.
		
		Siwei Tu (tusiwei@hunnu.edu.cn) is with School of Information Science and Engineering, Hunan Normal University, Changsha, China.
		
		Li Dong (Dlj2017@hunnu.edu.cn) is with Changsha Social Laboratory of Artificial Intelligence, Hunan University of Technology and Business, Changsha, China.
		
		%Kezhi Wang (Kezhi.Wang@brunel.ac.uk) is with the Department of Computer Science, Brunel University London, UK.
		
		%Kun Yang (kunyang@essex.ac.uk) is with the School of Computer Science and Electronic Engineering, University of Essex, Colchester, CO4 3SQ, U.K., also with Changchun Institute of Technology.
		
		Cunhua Pan (cpan@seu.edu.cn) is with the National Mobile Communications Research Laboratory, Southeast University, Nanjing 210096, China.
		
		Jiangzhou Wang (j.z.wang@seu.edu.cn) is with the National Mobile Communications Research Laboratory, Southeast University, Nanjing, China, and also with the Purple Mountain Laboratories, Nanjing, China.
		
		Xiaohu You (xhyu@seu.edu.cn) is with the National Mobile Communications Research Laboratory, Southeast University, Nanjing, China, and also with the Purple Mountain Laboratories, Nanjing, China.
	}
	%	\thanks{The paper was submitted on \today.}\\
}

\markboth{Submitted for Review}%
{Shell \MakeLowercase{\textit{et al.}}: Bare Demo of IEEEtran.cls for IEEE Journals}



\maketitle


\begin{abstract}
The rapid development of generative Artificial Intelligence (AI) continually unveils the potential of Semantic Communication (SemCom). 
However, current talking-face SemCom systems still encounter challenges such as low bandwidth utilization, semantic ambiguity, and diminished Quality of Experience (QoE).
This study introduces a Large Generative Model-assisted Talking-face Semantic Communication (LGM-TSC) System tailored for the talking-face video communication. 
 Firstly, we introduce a Generative Semantic Extractor (GSE) at the transmitter based on the FunASR model to convert semantically sparse talking-face videos into texts with high information density. 
 Secondly, we establish a private Knowledge Base (KB) based on the Large Language Model (LLM) for semantic disambiguation and correction, complemented by a joint knowledge base-semantic-channel coding scheme.
Finally, at the receiver, we propose a Generative Semantic Reconstructor (GSR) that utilizes BERT-VITS2 and SadTalker models to transform text back into a high-QoE talking-face video matching the user's timbre.
Simulation results demonstrate the feasibility and effectiveness of the proposed LGM-TSC system.
\end{abstract}


\section{Introduction}
% In the future, 6G will blur the boundaries between reality and virtuality, reshaping our world to accommodate diverse communication entities, including humans, machines, objects, and spirits. The information exchange between these entities will require higher intelligence, precision, and simplicity \cite{uusitalo20216g}. Therefore, the 6G network will have to facilitate the wireless transmission of huge data volumes, insist on swift system responses, and ensure trustworthy and efficient information interaction \cite{yang2022semantic}.
In the future, 6G will revolutionize communication and significantly enhance the flow of information and the sharing of knowledge. However, a substantial portion of the global population still lacks access to reliable high-speed network connections, or experiences instability when using them \cite{tandon2022txt2vid}. Addressing these challenges requires the development of technologies capable of efficiently transmitting video under ultra-low bandwidth conditions. Such advancements will not only boost network efficiency and mitigate issues caused by instability but also expand access to information and knowledge in regions without high-speed networks, thereby effectively narrowing the global digital divide.

In video communications, the most common type is the talking-face video, where facial expressions and mouth movements are synchronized with the audio \cite{chen2023compact}. Researchers have developed various methods to enhance the quality and efficiency of talking-face video transmission \cite{zhang2020davd}. Among these approaches, Semantic Communication (SemCom) emerges as a novel communication paradigm that seeks to enable smarter and more efficient transmission by comprehending and leveraging semantic content. For instance, some SemCom systems have achieved compression of talking-face videos by modeling human faces across successive frames \cite{du2023talking}. However, existing talking-face communication methods still encounter the following challenges:

\begin{enumerate}[]
	\item {\it{Low bandwidth utilization:}}
Video typically consists of a large number of frames containing substantial pixel information. Traditional video communication systems encode videos at the pixel level, resulting in inefficient bandwidth utilization. Given the continuity and similarity of video content, there is often significant redundancy in information between adjacent pixels and frames, leading to low semantic and information density. Consequently, effectively compressing video while enhancing the efficiency of information transmission without losing critical semantics is an urgent challenge that needs to be addressed \cite{yeo2018neural}.

	\item {\it{Semantic ambiguity:}}
	The raw data from the transmitter may be ambiguous for various reasons. 
	First, users may employ vague or ambiguous language, leading to unclear intentions in the message. Even when the user's data is precise, the receiver might interpret the information differently from how it was meant to be understood due to suboptimal channel conditions during transmission. Therefore, it is crucial to reduce ambiguity and errors throughout the communication process and ensure accurate information transmission for enhancing communication quality \cite{giorgetti2005effect}.
	
	\item {\it{Diminished Quality of Experience (QoE):}} 
	Due to lossy compression, the direct transmission of talking-face videos using previous deep-learning-based SemCom systems may result in noticeable differences in timbre and facial movements compared to the source video, significantly reducing the QoE for users. Thus, maintaining a high QoE for users during talking-face video transmission in SemCom systems is of paramount importance \cite{li2022cross}.
\end{enumerate}

With the rapid advancement of generative Artificial Intelligence (AI), Large Generative Models (LGMs) have become increasingly adept at producing diverse and rich content across various modalities, including text, images, audio, and video. Models such as GPT-4 \cite{achiam2023gpt}, Gemini Pro \cite{team2023gemini}, and LLaVA \cite{liu2024visual} have achieved cross-modal information processing and content generation through extensive data learning.
By leveraging their robust capabilities in semantic understanding and generation, these LGMs offer significant advantages in Semantic Communication (SemCom), including precise semantic extraction, extensive prior knowledge, and effective semantic disambiguation \cite{guo2024semantic}. These strengths provide novel insights and solutions for achieving efficient and intelligent SemCom.
This paper introduces a Large Generative Model-assisted Talking-face Semantic Communication System (LGM-TSC), built upon the capabilities of LGMs. Specifically, our primary contributions include:
\begin{enumerate}[]
	\item {\it{Generative Semantic Extractor (GSE):}}
We propose a GSE at the transmitter that converts talking-face video into text by utilizing an advanced Automatic Speech Recognition (ASR) model for high-precision speech recognition. By transforming the continuous audio signals in the video into discrete text, we achieve efficient compression of the talking-face video. This process significantly reduces the required bandwidth for data transmission while accurately extracting the key semantic content from the video.
	
	\item {\it{Large Language Model (LLM)-based private Knowledge Base (KB):}}
To address the potential semantic ambiguity and signal interference during SemCom, we introduce a private KB powered by LLM. 
Leveraging the robust semantic understanding and generative capabilities of LLMs, this private KB is employed for semantic disambiguation and text correction at both ends of the SemCom system. 
Furthermore, we have developed a joint knowledge base-semantic-channel coding method to further optimize the performance of the SemCom system, which can enable the LLM to indirectly participate in joint encoding.

	\item {\it{Generative Semantic Reconstructor (GSR):}}
 At the receiver, we propose a GSR that accurately converts text back to high-fidelity talking-face video. This is accomplished using BERT-VITS2 and SadTalker models to generate high-quality audio and video, thereby ensuring an enhanced QoE for users. By leveraging the user's face image and vocal feature vector stored in the public KB, the GSR is capable of restoring the dynamic realism of the original talking-face video while maintaining consistency with the user's vocal timbre, providing users with an immersive viewing experience.
\end{enumerate}

The rest of this paper is structured as follows. Section II presents the related work, Section III introduces the system model, Section IV provides a detailed description of the proposed LGM-TSC system, Section V outlines the experimental setup and results, and Section VI concludes the paper.

\section{Related Work}
% This section introduces the preliminaries about the key GAMs used in this paper, including NeRF, SAM, CGAN, and DM.

\subsection{Deep learning-based video SemCom systems}
Gao et al. \cite{gao2024semantic} proposed an Adaptive Panoramic Video Semantic Transmission (APVST) network based on a Deep Joint Source-Channel Coding (DeepJSemComC) architecture and attention mechanisms, which significantly reduces channel bandwidth costs compared to other SemCom and traditional video transmission schemes. Li et al. \cite{li2024video} introduced an end-to-end video SemCom system that employs Main Object Extraction (MOE) and Contextual Video Encoding (CVE) techniques to achieve efficient video transmission.
Zhang et al. \cite{zhang2023deep} proposed a deep learning-based video transmission SemCom system that uses dual optical flow to estimate inter-frame residual details and employs feature selection, feature fusion, and noise attention modules to enhance transmission efficiency and system robustness. %This approach reduces the number of transmitted symbols by approximately 33.3\% while increasing the PSNR performance by an average of 0.56 dB. 
Tandon et al. \cite{tandon2022txt2vid} introduced a novel video compression pipeline named Txt2Vid, which compresses camera video into text transcripts and then decodes the text back into realistic reconstructions of the original video using deep learning-based speech generation and lip synchronization models. %This method achieves bitrate reductions of two to three orders of magnitude compared to standard audio-video codecs.

\subsection{Joint coding in SemCom systems}
Xu et al. \cite{xu2023deep} explored the application of DeepJSemComC in SemCom, presented an adaptive DeepJSemComC architecture, and demonstrated its superior performance and flexibility in various tasks such as image and video transmission, channel state information feedback, image retrieval, and multi-agent communication. %It also discusses future research directions regarding the security of DeepJSemComC, the development of universal encoder/decoder architectures for multi-modal data sources, and challenges to be addressed before its practical implementation. 
Tung et al. \cite{tung2023deep} introduced Deep joint source-channel and encryption coding (DeepJSemComEC), a scheme for wireless image transmission, which achieves secure communication against eavesdroppers without relying on assumptions about their channel quality or the intended use of the intercepted signals. %Moreover, it demonstrates that the proposed encryption method is applicable to other end-to-end joint source-channel coding problems, such as remote classification, without modification, while also providing graceful degradation of image quality in varying channel conditions. 
Dai et al. \cite{dai2022nonlinear} introduced a novel class of high-efficiency deep joint source-channel coding methods known as Nonlinear Transform Source-Channel Coding (NTSemComC), which adapted closely to the source distribution through nonlinear transformation and optimizes end-to-end transmission performance by learning latent representations and entropy models of the source data. %thereby supporting the demands of the future SemCom.

\subsection{LLM for SemCom systems}
Guo et al. \cite{guo2024semantic} proposed a communication system that was aware of semantic importance, leveraging large pre-trained language models for semantic understanding and error correction of visual data. This system converts visual content into natural language descriptions to enhance semantic comprehension during transmission while minimizing semantic loss under transmission delay constraints. %Additionally, the method improves privacy protection by not directly transmitting raw data or features. 
Jiang et al. \cite{jiang2023large} introduced a multimodal SemCom framework based on large AI models. This framework employs multimodal language models for data alignment, a personalized LLM knowledge base for semantic extraction and recovery, and conditional generative adversarial networks for channel estimation, thereby enhancing signal transmission quality in fading channels. Furthermore, Park et al. \cite{park2023towards} explored the SemCom protocols in 6G systems, proposing a three-tier classification from task-oriented neural protocols to semantic protocols based on LLM. It further discussed how these protocols leverage LLM to adapt to diverse and dynamic task requirements.

However, current research rarely explores the simultaneous use of visual generative models and LLMs to enhance the performance of SemCom systems. Moreover, existing methods do not account for the joint encoding of LLMs with semantic and channel encoders. Our study thoroughly investigates the application of visual generative models to improve the QoE in SemCom systems and incorporates the joint encoding of LLMs, semantic encoders, and channel encoders. As a result, our approach achieves superior performance compared to previous works.

\section{System Model}

\begin{figure}[htbp]
	\centering
	\includegraphics[width=8.5cm]{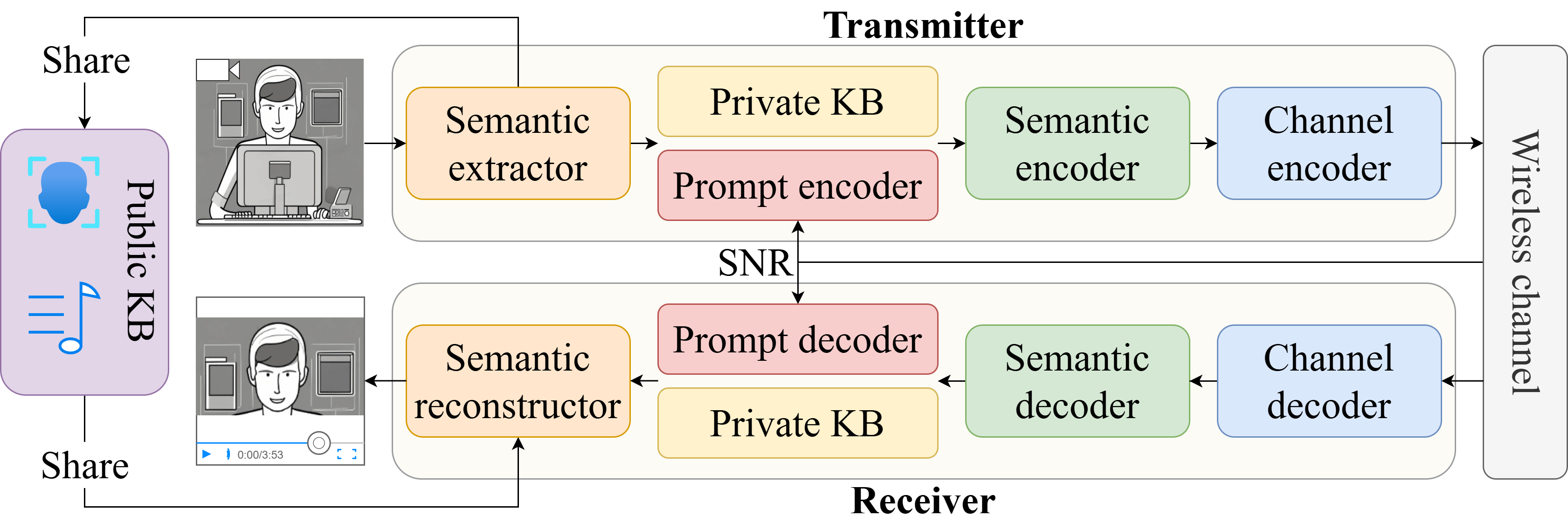}
	\caption{The system model of the proposed SemCom system.}
	\label{fig:system}
\end{figure}

The proposed LGM-TSC system comprises four main components: a transmitter, a receiver, a physical channel, and a public KB, as depicted in Fig. \ref{fig:system}.

\subsection{Transmitter}
The input to the transmitter is matrix \(\mathbf{X} \in \mathbb{R}^{F \times H \times W \times C}\), which represents the talking-face video and has a size of $\mathit{F}$ (Number of frames) × $\mathit{H}$ (height) × $\mathit{W}$ (width) × $\mathit{C}$ (channel).
At the transmitter, the input talking-face video \textbf{X} is mapped to symbol \textbf{Y} for transmission over a physical channel. 
The transmitter comprises four components: a GSE for semantic extraction, a private KB based on an LLM with a prompt encoder, a semantic encoder, and a channel encoder. 
The GSE converts talking-face video into semantically aligned text. 
The private KB disambiguates the text and performs the knowledge base encoding using  the prompt encoder. 
The semantic encoder and channel encoder are responsible for semantic encoding and channel encoding, respectively, ensuring that the encoded semantics are transmitted smoothly over the physical channel.
The encoded symbol sequence \textbf{Y} can be expressed as:

\begin{equation}\label{eq:Shi1}
\mathbf{Y}=C_\beta\left(S_\alpha(K\left(p_\eta(\mu),E_\varepsilon\left(\mathbf{X}\right)\right))\right)
\end{equation}
where $E_\varepsilon\left(\cdot\right)$ is the GSE with the parameter set $\varepsilon$, $K\left(\cdot\right)$ is the private KB based on the LLM, $p_\eta\left(\cdot\right)$ is the prompt encoder of the private KB whose parameter set is $\eta$, $\mu$ is the Signal-to-Noise Ratio (SNR) parameter of the channel, $S_\alpha\left(\cdot\right)$ is the semantic encoder with the parameter set $\alpha$, and $C_\beta\left(\cdot\right)$ is the channel encoder with the parameter set $\beta$.

\subsection{Wireless channel}
The transmitter sends encoded symbols \textbf{Y}, which are transmitted through the physical channel to the receiver. The channel output sequence $\hat{\mathbf{Y}}$ at the receiver can be expressed as:
\begin{equation}\label{eq:Shi1}
\hat{\mathbf{Y}}=\ \mathbf{HY}\ +\ \mathbf{n}
\end{equation}
where \textbf{H} represents the channel gain, and \textbf{n} is the Additive White Gaussian Noise (AWGN).

\subsection{Public KB}
\textcolor{black}{
	We set up a public KB that interacts with the transmitter and receiver. 
	The transmitter verifies the user's face image \textcolor{black}{\(\textbf{P}\)} and vocal feature vector \(\textbf{v}\) and stores them in advance in the public KB. 
	At the receiver, GSR will use these features for reconstruction to ensure a high QoE talking-face video.}

\subsection{Receiver}
Similar to the transmitter, the receiver is composed of four components: a channel decoder, a semantic decoder, a private KB based on an LLM with a prompt decoder, and a GSR for semantic reconstruction. 
The semantic decoder and channel decoder are responsible for decoding text from received symbols. 
The private KB corrects any potential errors in the text and performs knowledge base decoding with the prompt decoder. 
The GSR then transforms the text into audio and subsequently into talking-face video. The reconstructed video can be represented as:
\begin{equation}\label{eq:Shi1}
\hat{\mathbf{X}}=R_\zeta(K(p_\eta^{-1}(\mu),S_\delta^{-1}(C_\gamma^{-1}(\hat{\mathbf{Y}}))),\textbf{P},\textbf{v})
\end{equation}
where $C_\gamma^{-1}\left(\cdot\right)$ is the channel encoder with the parameter set $\gamma$, $S_\delta^{-1}\left(\cdot\right)$ is the semantic encoder with the parameter set $\delta$, $K\left(\cdot\right)$ is the private KB based on the LLM, $p_\eta^{-1}\left(\cdot\right)$ is the prompt decoder of the private KB whose parameter set is $\eta$, $\mu$ is the SNR parameter of the channel, and $R_\zeta\left(\cdot\right)$ is the GSR with the parameter set $\zeta$.

To reconstruct audio and video information from the semantic level, it is crucial to maintain the consistency of textual semantics between \textbf{T} and $\hat{\mathbf{T}}$, where \(\mathbf{T}=E_\varepsilon\left(\mathbf{X}\right)\) represents the semantic text extracted by the GSE from the talking-face video at the transmitter, and $\hat{\mathbf{T}}=K(p_\eta^{-1}(\mu),S_\delta^{-1}(C_\gamma^{-1}(\hat{\mathbf{Y}})))$ represents the recovered semantic text after decoding. We utilize Cross Entropy (CE) as the loss function:
% \begin{equation}\label{eq:Shi1}
% L_{CE}\left(\mathbf{T},\hat{\mathbf{T}}\right)=-\sum_{l=1}^{L}{q\left(w_l\right)log{\left(p\left(w_i\right)\right)}+\left(1-q\left(w_l\right)\right)log{\left(1-p\left(w_i\right)\right)}},
% \end{equation}
\begin{equation}\label{eq:ShiCE}
\begin{split}
L_{CE}\left(\mathbf{T},\hat{\mathbf{T}}\right)=&-\sum_{l=1}^{L} \big[ q\left(w_l\right) \log{\left(p\left(w_i\right)\right)} \\
&+ \left(1-q\left(w_l\right)\right) \log{\left(1-p\left(w_i\right)\right)} \big]
\end{split}
\end{equation}
where $q\left(w_l\right)$ denotes the real probability of the appearance of the $\mathit{l}$-th word $w_l$ in $\mathbf{T}$, and $p\left(w_l\right)$ represents the predicted probability of the appearance of the $\mathit{l}$-th word $w_i$ in $\hat{\mathbf{T}}$. CE is employed to measure the difference between two probability distributions. By minimizing the CE loss, the semantic encoder and decoder can learn the word distribution $q\left(w_l\right)$ in $\mathbf{T}$, which represents the meaning of words in terms of grammar, phrases, and contextual information. Hence, the goal of the SemCom system is to determine the parameters of semantic encoder/decoder, prompt encoder/decoder and channel encoder/decoder $\alpha^\ast,\ \beta^\ast,\ \eta^\ast,\ \gamma^\ast\ \mathrm{and} \ \delta^\ast$ that minimize the expected distortion as follows:
\begin{equation}\label{eq:Shi1}
\left(\alpha^{*}, \beta^{*}, \eta^{*}, \gamma^{*}, \delta^{*}\right)=\arg \min _{\alpha, \beta, \eta, \gamma, \delta} E_{P(\mu)} E_{P(\mathbf{T}, \hat{\mathbf{T}})}\left[L_{C E}(\mathbf{T}, \hat{\mathbf{T}})\right]
\end{equation}
where $\alpha^\ast$ is the optimal semantic encoder parameters, $\beta^\ast$ is the optimal channel encoder parameters, $\eta^\ast$ is the optimal prompt encoder and decoder parameters, $\gamma^\ast$ is the optimal channel decoder parameters, and $\delta^\ast$ is the optimal semantic decoder parameters. \( P(\mathbf{T}, \hat{\mathbf{T}}) \) represents the joint probability distribution of $\mathbf{T}$ and $\hat{\mathbf{T}}$, and \( P(\mu) \) represents the probability distribution of the SNR.

\section{Proposed LGM-TSC System}
\begin{figure*}[htbp]
	\centering
	\includegraphics[width=18cm]{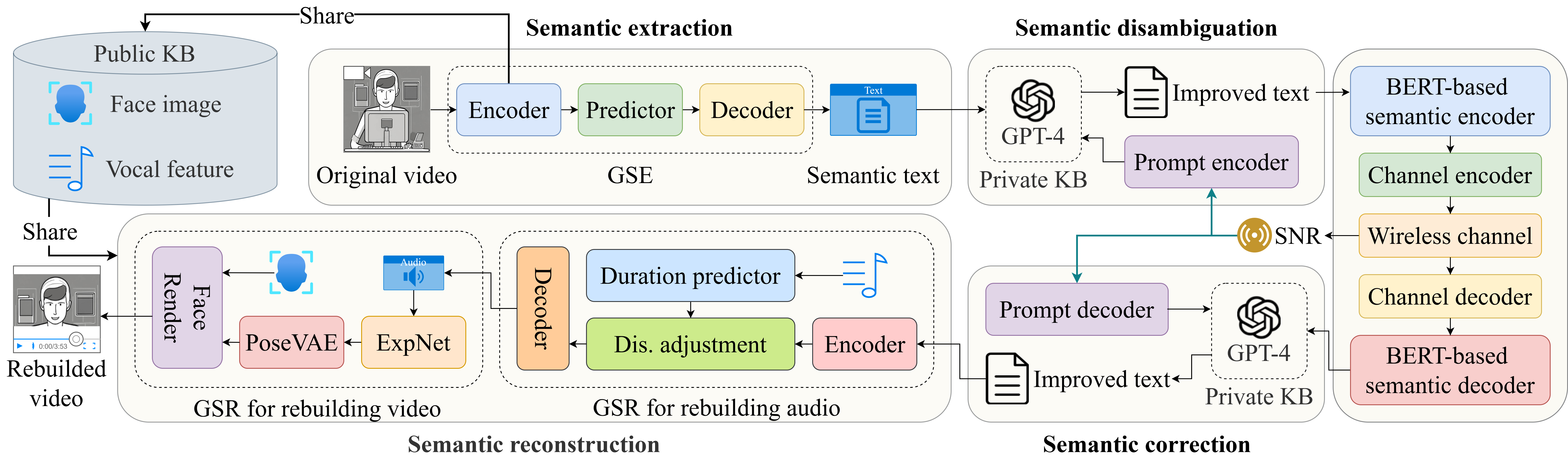}
	\caption{The proposed LGM-TSC system.}
	\label{fig:UBCCSemCom}
\end{figure*}

In this section, we provide the implementation details of the proposed LGM-TSC system, as illustrated in Fig. \ref{fig:UBCCSemCom}.

\subsubsection{Semantic extraction}
To mitigate the high bandwidth consumption associated with talking-face video transmission, we employ the FunASR model \cite{gao2023funasr} to construct the GSE at the transmitter. The FunASR model efficiently processes audio spectrum features and timbre. By leveraging this model to compress the original video into text, we transform the video stream, which requires substantial bandwidth, into text with minimal transmission demands, thereby achieving ultra-low bitrate compression of the talking-face video.

\subsubsection{Semantic disambiguation and correction}
To address semantic ambiguity and errors in SemCom, we introduce a private KB based on an LLM for disambiguation and correction. At the transmitter, the private KB performs preliminary semantic analysis of the text extracted by the GSE. It identifies potential ambiguities based on background knowledge and contextual information, resolves these ambiguities, and then dynamically perform knowledge base encoding according to SNR. 

At the receiver, the private KB corrects the received text based on contextual information and performs knowledge base decoding using keywords in the text to generate a more accurate and coherent semantic representation. We have developed a prompt encoder and decoder, which function as a fuzzy inference system that adaptively controls the length of the LLM's output content based on varying SNR levels. This enables the realization of joint knowledge base-semantic-channel coding.

\subsubsection{BERT-based semantic encoder and decoder}
The text processed by the private KB is input into a BERT-based semantic encoder \cite{devlin2018bert}, which comprises multiple transformer encoder layers \cite{vaswani2017attention}. At the receiver, the semantic decoder is implemented as the final linear layer of the BERT model.

\subsubsection{Channel encoder and decoder}
The encoded semantic features are passed through the channel encoder to ensure effective transmission of semantics over the physical channel. The channel encoder consists of an autoencoder \cite{ng2011sparse}, with the feature dimension of the hidden layers progressively decreasing to achieve data compression. To maintain information consistency, the channel decoder employs an autoencoder structure that mirrors that of the channel encoder.

\subsubsection{Semantic reconstruction}
At the receiver, semantic reconstruction is critical for successful SemCom. We utilize GSR to convert the text back into the original talking-face video. GSR comprises two key components: the BERT-VITS2 model for audio reconstruction and the SadTalker model \cite{zhang2023sadtalker} for video reconstruction. The BERT-VITS2 model first processes the text from the semantic decoder to reconstruct the audio, relying on the user's vocal feature vector stored in the public KB to ensure timbre similarity and realism. The reconstructed audio is then fed into the SadTalker model, which uses both the audio and the user's face image stored in the public KB to generate realistic talking-face videos.
%The SadTalker model employs advanced 3D motion coefficient learning methods to precisely control expressions and head postures of individuals in the video, ensuring that generated videos are visually consistent with their original counterparts.

%\subsubsection{ Training process of UBC-CSemCom system}
%It is important to note that the GSE, based on FunASR, and GSR, based on BERT-VITS2 and Sadtalker, are pre-training models and do not participate in the training of the CSemCom system. We design a prompt encoder and decoder based on neural fuzzy inference system for knowledge base encoding and train it with semantic/channel encoder and decoder.

Next, we provide a detailed explanation of each contribution presented in this paper. The workflow of the proposed LGM-TSC system is illustrated in \textbf{Algorithm} \ref{algall}.

\begin{algorithm}
	\caption{LGM-TSC system}
	\label{algall}
	\begin{algorithmic}[1]
		\REQUIRE Original talking-face video \(\textbf{X}\)
		\ENSURE Reconstructed talking-face video \(\hat{\textbf{X}}\)
		\STATE{Use GSE to compress \(\textbf{X}\) and extract the semantic text $\mathbf{T}$ according to \textbf{Algorithm} \ref{alg1}.}
		\STATE{Perform the joint knowledge base-semantic-channel encoding at the transmitter.}
		\STATE{Transmit semantic features over the physical channel.}
        \STATE{Perform the joint knowledge base-semantic-channel decoding, and obtain the recovered semantic text  $\hat{\mathbf{T}}$ from the private KB.}
        \STATE{Use GSR to perform semantic reconstruction to obtain $\hat{\mathbf{X}}$ according to \textbf{Algorithm} \ref{alg2}.}
		\STATE{Return $\hat{\mathbf{X}}$.}
	\end{algorithmic}
\end{algorithm}

\subsection{FunASR-based GSE for semantic extraction}
FunASR is a sophisticated ASR model designed to develop high-precision audio speech recognition services \cite{gao2023funasr}, and it is an end-to-end Non-AutoRegressive (NAR) model, resulting in faster inference speed. At the transmitter, the GSE is constructed using the FunASR model, and audio in the talking-face video is converted into corresponding text using three modules: Encoder, Predictor, and Decoder. The workflow of the FunASR-based GSE is illustrated in Fig. \ref{fig:asr}. 
The process of extracting semantics from talking-face video \textbf{X} at the transmitter, and generating semantic text \textbf{T} is as follows:

\begin{figure}[htbp]
	\centering
	\includegraphics[width=8.5cm]{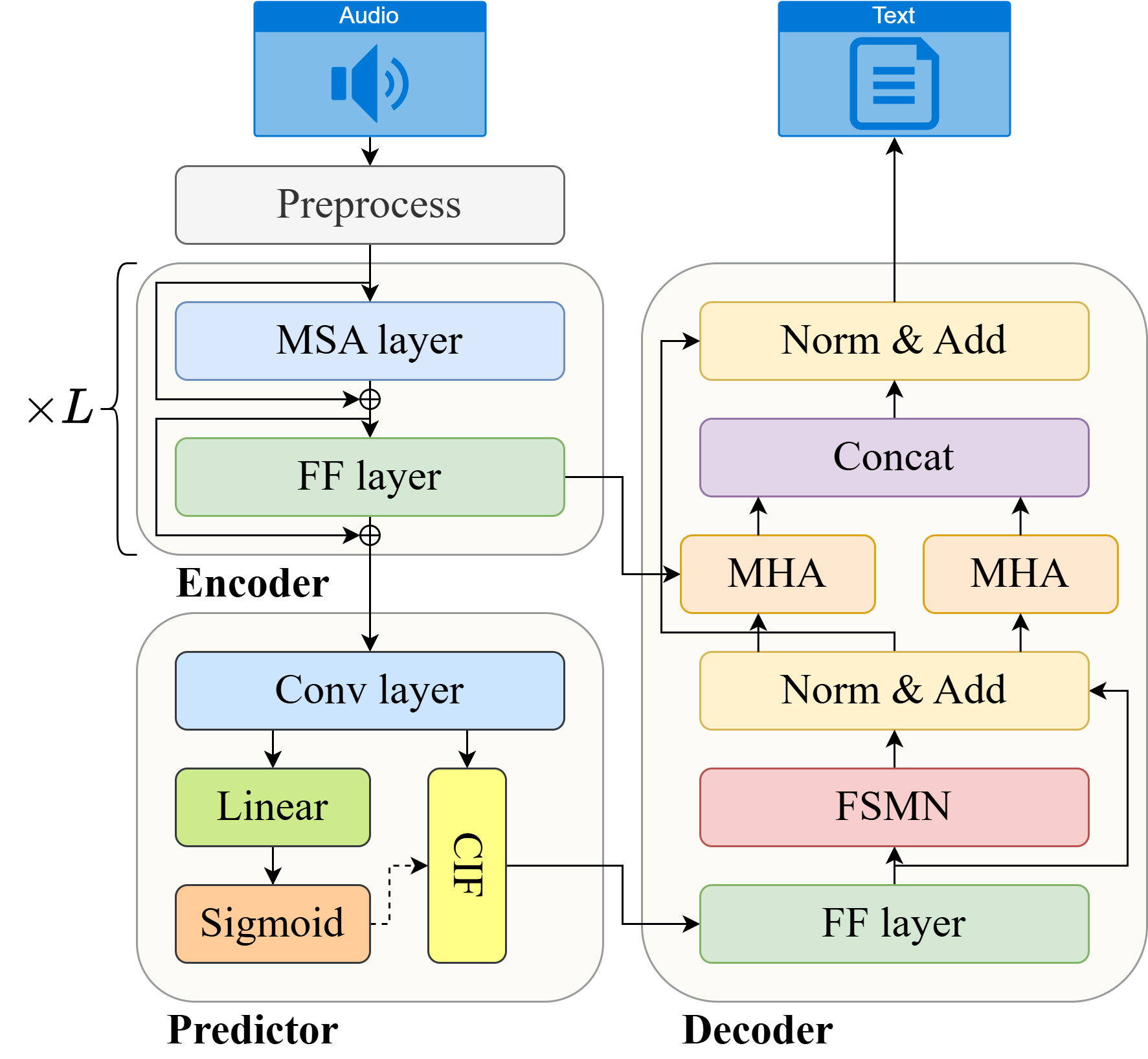}
	\caption{The architecture of the GSE.}
	\label{fig:asr}
\end{figure}

\subsubsection{Encoder}
Audio is a continuous signal that cannot be directly understood by the model. Therefore, it is necessary to carry out discrete sampling to obtain the feature vector. 
\textcolor{black}{
	Mel Filter Bank (FBank) is a common speech feature. The process of extracting Fbank from audio is similar to the process of human ear processing audio, which can improve the performance of speech recognition \cite{2014Research}.}
We first extract FBank feature \textbf{F} from video \textbf{X}. 
	Subsequently, the Encoder with multiple layers which contain the Multi-head Self Attention (MSA) layer and Feed Forward (FF) layer will extract high-dimensional features from \textbf{F}. 
	The specific workflow of the Encoder is as follows:
	
\begin{itemize} 
	\item First, the MSA layer allows the vector of
	each time step to interact with vectors of all other time steps,
	capturing both global and local information in audio.
	The output of the MSA layer can be calculated as follows:
	\begin{equation}
		\textcolor{black}{
			\mathbf{M}_{msa,i} = \left\{
			\begin{array}{l}
				\mathrm{Norm}(\mathrm{MSA}(\mathbf{F}) + \mathbf{F}), \, i = 1 \\
				\mathrm{Norm}(\mathrm{MSA}(\mathbf{M}_{ff,i-1}) + \mathbf{M}_{ff,i-1}), \, i \ge 2
			\end{array}
			\right.
		}
		\label{eq:gs6}
	\end{equation}
	
	where $\mathrm{MSA}\left(\cdot\right)$ is the MSA operator and $\mathrm{Norm}\left(\cdot\right)$ is the layer normalization operator.
	
	\item The features output by the MSA layer are then input to the FF layer. The FF layer comprises a linear layer and an activation function that applies a nonlinear transformation to the vector for each time step, thereby enhancing the adaptability of the model. The output of the FF layer in the first encoder layer is:
	\begin{equation}
		\mathbf{M}_{ff,i} = \mathrm{Norm}\left (\text{GeLU}(\mathbf{W}_{b,f} \cdot  \mathbf{M}_{msa,i}  + \mathbf{b}_{b,f}) + \mathbf{M}_{msa,i}  \right ) 
		\label{eq:gs7}
	\end{equation}
	
	where $\mathbf{W}_{b,f}$ and $\mathbf{b}_{b,f\ }$ are the weights and biases of the FF layer in the encoder layer, and GeLU\(\left(\cdot \right) \) denotes the activation function.
	
\end{itemize} 

Finally, the output of the Encoder with $L$ layers is:

\begin{equation}
	\mathbf{M}_L=\mathrm{Norm}\left(\mathbf{M}_{ff,L}\right)
	\label{eq:gs8}
\end{equation}
where $\mathbf{M}_{ff,L}$ means the output of the $\mathit{L}$-th layer of the Encoder.

\subsubsection{Predictor}
\textcolor{black}{
	As we all know, the duration of each word in the audio may not be only one frame, and the discrete FBank features are obtained by processing each frame as input to the model. 
	If the output of the Encoder is processed directly, it will inevitably lead to a large number of repeated characters. Therefore, we continue to use the Predictor module to extract acoustic hidden variables that are used to determine the length of the resulting text.
	The workflow for Predictor is as follows:}
	\begin{itemize}
	\item \textcolor{black}{
		Predictor starts by generating a weight for all vectors in $\mathbf{m}_{ff,L}$ that ranges from 0 to 1:
		\begin{equation}
			\alpha _{weight}=\mathrm{Softmax} \left ( \mathrm{Linear} \left ( \mathrm{Conv}\left ( \mathbf{M}_{ff,L} \right )  \right )  \right ) 
			\label{eq:pre}
		\end{equation}
		where \(\mathrm{Conv}\left(\cdot \right) \) is the convolution layer, \(\mathrm{Linear}\left(\cdot \right) \) and \(\mathrm{Softmax}\left(\cdot \right) \) are linear projection operator and softmax activation operator, respectively. \(\alpha _{weight}\) is the set of weights for each vector.}	

		\item \textcolor{black}{
			To aggregate the features corresponding to a single character over different times into independent acoustic latent variables, the Predictor draws inspiration from the principle of neurons accumulating charge until their membrane potential reaches a threshold and triggers an action potential. 			
			It employs a module named Continuous Integrate-and-Fire (CIF) \cite{dong2020cif}, which sequentially extracts an acoustic latent variable from a series of vectors with weights (i.e., $\alpha _{weight}$) summing to one, following the sequence order of the feature vectors. The number of acoustic latent variables is equal to the length of the resulting text, and we refer to all acoustic latent variables as $\mathbf{E}_a$.}
	\end{itemize}

\begin{comment}
	Based on the output of the Encoder module, the Predictor module generates acoustic hidden variables via the internal CIF module \cite{dong2020cif} that capture more detailed acoustic information, primarily for generating the final speech recognition result.
	
	In Fig, Token represents a single vector of $\mathbf{m}_L$ in the sequence length dimension. Token weight is assigned by the Predictor module for each token, which are generated through a convolutional layer, linear layer, and sigmoid activation function with a range of 0 to 1. $\mathbf{E}_a$ is the acoustic hidden variable calculated by the CIF module \cite{dong2020cif}. The weights of the tokens are accumulated in sequence, and when their sum reaches 1, the CIF module extracts an acoustic hidden variable vector $\mathbf{E}_{a,i}$ from these tokens. This process is repeated, and any weights exceeding 1 are carried over to the next calculation. Ultimately, the Predictor module outputs an acoustic hidden variable matrix:
	\begin{equation}\label{eq:gs9}
		\mathbf{E}_a=\mathrm{Concat}\left(\mathbf{E}_{a,1},\mathbf{E}_{a,2},\ldots\mathbf{E}_{a,l}\right)
	\end{equation}
	where $\mathbf{E}_{a,i}$ represents the $\mathit{i}$-th acoustic hidden variable synthesized, $\mathrm{Concat}\left(\cdot\right)$  denotes the vector concatenation operation, and the sequence length $\mathit{l}$ is equivalent to the number of words in the translated text....
\end{comment}

\subsubsection{Decoder}
The Decoder module decodes the final text based on the feature $\mathbf{M}_L$ extracted by the Encoder, the acoustic hidden variable $\mathbf{E}_a$ extracted by the Predictor, and the pre-set hot words, which can enable the FunASR model to focus more on key semantics when transcribing audio.

Pre-set hot words are transformed into vectors through embedding layers, representing $\mathbf{E}_h$. This allows $\mathbf{E}_h$ to capture the semantic features of hot words, enabling the model to more accurately identify and process these words in its internal representation. Subsequently, the Decoder module adjusts the acoustic hidden variable $\mathbf{E}_a$ to the appropriate dimension and calculates Multi-head Attention (MHA) with feature $\mathbf{M}_L$ and embedded hot word $\mathbf{E}_h$ respectively. Following a series of post-processing steps, the semantic text \textbf{T} is obtained:
\begin{equation}\label{eq:gs10}
{\mathbf{E}_a}^\prime=\mathrm{Norm}\left(\mathrm{FSMN}\left(\mathrm{FF}\left(\mathbf{E}_a\right)\right)\right)+\mathbf{E}_a
\end{equation}
\begin{equation}\label{eq:gs11}
\mathbf{T}=\mathrm{Norm}\left(\mathrm{Concat}\left(\mathrm{MHA}\left({\mathbf{E}_a}^\prime,\mathbf{M}_L\right),\\
\mathrm{MHA}\left({\mathbf{E}_a}^\prime,\mathbf{E}_h\right)\right)\right)+{\mathbf{E}_a}^\prime
\end{equation}
where ${\mathbf{E}_a}^\prime$ is an acoustic hidden variable of the same dimension as $\mathbf{M}_L$, $\mathrm{FF}\left(\cdot\right)$ is FF layer, \textcolor{black}{$\mathrm{FSMN}\left(\cdot\right)$ is Feedforward Sequential Memory Network (FSMN) which is used to detect the beginning and end of valid audio, thereby filtering out invalid audio parts \cite{zhang2015feedforward}}, $\mathrm{Norm}\left(\cdot\right)$ is the layer normalization operator, $\mathrm{MHA}\left(\cdot\right)$ is the MHA operator, and $\mathrm{Concat}\left(\cdot\right)$ denotes the vector concatenation operation.

The workflow of GSE is illustrated in \textbf{Algorithm} \ref{alg1}.

\begin{algorithm}
	\caption{GSE}
	\label{alg1}
	\begin{algorithmic}[1]
		\REQUIRE Original talking-face video $\bm{\mathrm{X}}$
		\ENSURE Semantic text $\mathbf{T}$
		\STATE{Extract Fbank features from $\bm{\mathrm{X}}$, and obtain audio feature $\mathbf{M}_L$ through feature extraction according to Eqs. (\ref{eq:gs6})-(\ref{eq:gs8})}.
		\STATE{Extract acoustic latent variables $\mathbf{E}_a$ from $\mathbf{M}_L$ according to Eq. (\ref{eq:pre}).}
		\STATE{Decode $\mathbf{E}_a$ to obtain \textbf{T} according to Eqs. (\ref{eq:gs10})-(\ref{eq:gs11}).}
		\STATE{Return \textbf{T}.}
	\end{algorithmic}
\end{algorithm}

\subsection{Private KB for semantic disambiguation and correction}
In order to ensure the accuracy and reliability of information in the transmission process, we introduce a private KB based on an LLM that has powerful semantic understanding and generation capabilities. 
In LGM-TSC, we use the private KB to disambiguate and correct the text extracted from the talking-face video.

At the transmitter, the private KB conducts a preliminary semantic analysis of the text extracted by the GSE. 
This involves entity recognition, relationship extraction, and semantic role labeling to identify potential ambiguity points in the text and make initial corrections. 
Subsequently, the private KB uses the prompt encoder to perform knowledge base encoding on the text.

\begin{figure*}[htbp]
	\centering
	\includegraphics[width=17cm]{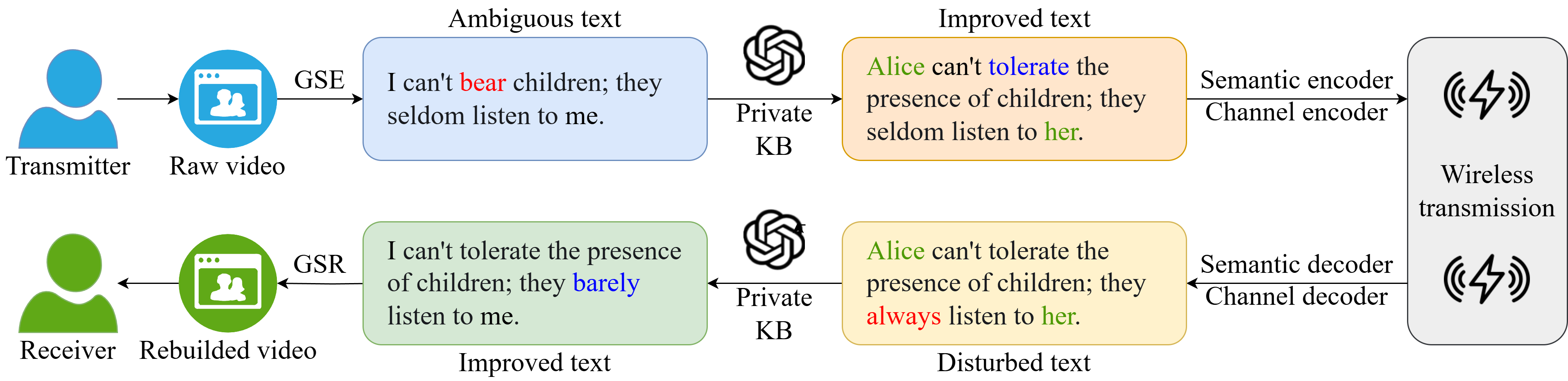}
	\caption{\textcolor{black}{Example of disambiguation and correction by the private KB.}}
	\label{fig:gkb1}
\end{figure*}

At the receiver, the decoded text may contain errors or new ambiguities due to factors such as channel noise. 
Once again, the private KB performs an in-depth analysis of the received text, utilizing its extensive prior knowledge and robust semantic understanding to rectify any inaccuracies in the text. 
Subsequently, the private KB uses the prompt decoder to perform knowledge base decoding on the text, resulting in a more comprehensive semantic expression of the text. Fig. \ref{fig:gkb1} provides an example of semantic disambiguation and correction.

\begin{figure*}[htbp]
	\centering
	\includegraphics[width=17cm]{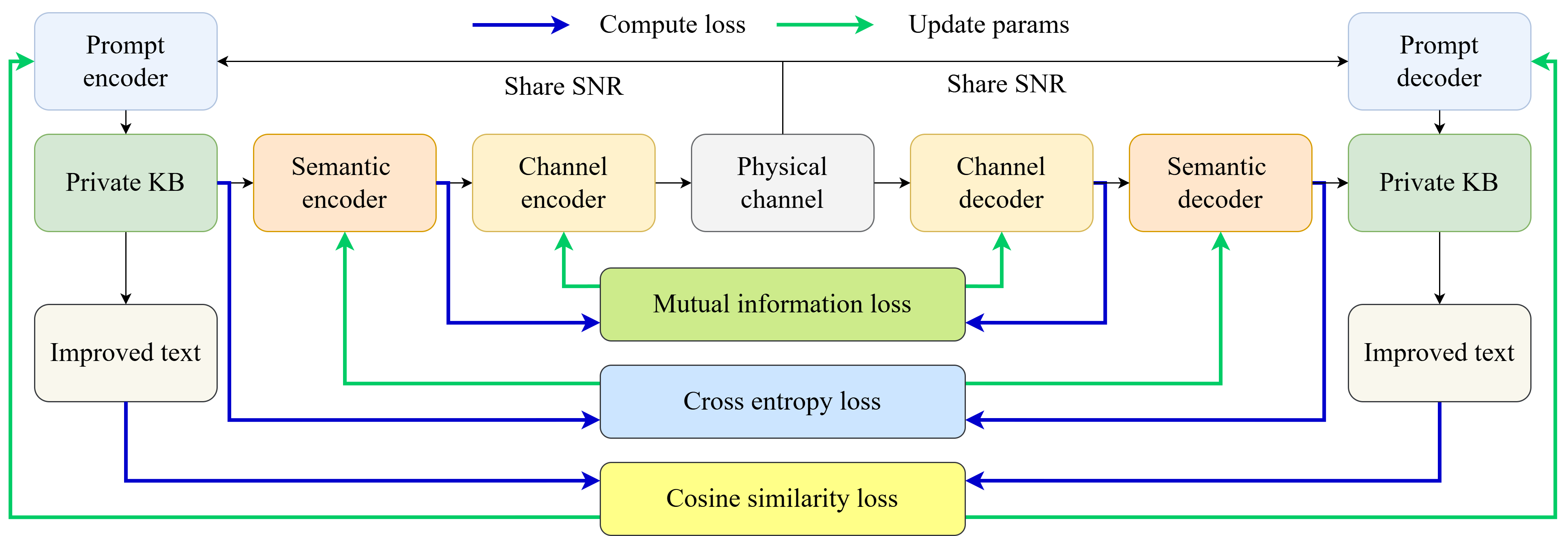}
	\caption{Framework of the joint knowledge base-semantic-channel coding.}
	\label{fig:gkb2}
\end{figure*}

\begin{comment}
	\begin{figure*}[htbp]
		\centering
		\includegraphics[width=17cm]{gkbtable.png}
		\caption{\textcolor{black}{Example of prompt encoding for the private KB.}}
		\label{fig:gkbtable}
	\end{figure*}
\end{comment}

\begin{table*}[]
	\centering
	\normalsize 
	\renewcommand{\arraystretch}{1.5}
	\caption{Example of prompt encoding for the private KB.}
	\label{tab:exam}
	\begin{tabular}{|l|l|}
		\hline
		%\rowcolor[HTML]{FFE6CC} 
		Fuzzy rules & Output text after knowledge base encoding \\ \hline
		%\rowcolor[HTML]{DAE8FC} 
		If SNR is low, then length = 70\%-80\%    & A young child, smiling, slides down the slide.                      \\ \hline
		%\rowcolor[HTML]{DAE8FC} 
		If SNR is middle, then length = 80\%-90\% & A young child, with a smile, eagerly slides down the slide.         \\ \hline
		%\rowcolor[HTML]{DAE8FC} 
		If SNR is high, then length = 100\%       & A young child, with a beaming smile, eagerly slides down the slide. \\ \hline
	\end{tabular}
	\vspace{0.25cm} % 可选：调整表格和附加文本之间的距离
	
	\centering
	In this example, we assume the input text for the LLM is ``A young child, with a beaming smile, eagerly slides down the slide." Meanwhile, the length in fuzzy rules means the output length of LLM when proceeding with knowledge base encoding in the left column of the table.
\end{table*}

\subsubsection{Disambiguation} Assuming that the text obtained after GSE compression of the talking-face video at the transmitter is, ``I can't bear children; they seldom listen to me." Generally, the private KB may identify two possible interpretations: the first being, ``I can't give birth to children," and the second being, \textcolor{black}{``I can't tolerate children."} At the same time, let us assume that the private KB already contains information indicating that the user is a woman named Alice who has given birth to a child as part of the background. As a result, the private KB judges that the user is more likely to express frustration with children rather than an inability to have them. Therefore, the private KB retains this intent and makes appropriate adjustments to the text to remove ambiguities. For example, it might change the text to, ``Alice can't tolerate the presence of children; they seldom listen to her." Such statements more clearly convey the user's actual intent.

\subsubsection{Correction} At the receiver, even if the text at the transmitter has been disambiguated, channel noise and other interference during transmission may still cause errors or ambiguities in the decoded text. The private KB will conduct an in-depth analysis of the received text and make secondary corrections based on known user settings. For example, if the decoded text at the receiver is, ``Alice can't tolerate the presence of children; they always listen to her," the private KB will identify this error based on background information about Alice's identity and correct it to ``I can't tolerate the presence of children; they barely listen to me." These corrections ensure the accuracy and completeness of the semantics, allowing the receiver to obtain information consistent with the user's original intent.

\vspace*{-3mm} 

\subsection{Joint knowledge base-semantic-channel coding}
To make the SemCom system better cope with the wireless environment, we design a novel joint coding scheme. 
Due to the large parameter size of LLM-based KB, the KB is challenging to participate in the joint encoding of SemCom systems. We designed a prompt encoder capable of controlling the output of the LLM, enabling the LLM to indirectly participate in joint encoding. This facilitates high-quality joint knowledge base-semantic-channel coding.

\begin{comment}
The LLM of GKB processes the text information based on the prompt encoder and decoder, resulting in dynamic changes to the semantic output of the text according to SNR. 	
\end{comment}

At the transmitter, we utilize the prompt encoder and the LLM-based private KB for performing knowledge base encoding on the disambiguated text. This is combined with semantic encoding and channel encoding to create a process called joint knowledge base-semantic-channel encoding.

%To prevent the occurrence of the ``cliff effect" resulting from sudden changes in channel conditions, 
We develop a prompt encoder and decoder based on a fuzzy inference system. 
The input of the fuzzy inference system is the SNR value of the channel, and the fuzzy semantic set is \{high, medium, low\}, which corresponds to the prompt encoding/decoding rules when SNR is different. When the SNR is high, the prompt encoder directs the LLM to produce more accurate text content. Conversely, when the SNR is low, the prompt encoder instructs the LLM to compress the output text while maintaining semantic integrity. The structure of the prompt encoder\cite{jang1993anfis} is as follows:

\subsubsection{Fuzzification layer}
This layer generates the corresponding fuzzy membership levels based on the input SNR. The output of this layer is:
\begin{equation}\label{eq:Shi1}
O_i^1=\mu_{A_i}\left(x\right),\forall i=1,2,3
\end{equation}
where $i$ corresponds to different rules; $x$ is the input; $A_i$ are fuzzy linguistic labels (representing high, medium, and low SNR); $\mu_{A_i}$ is the fuzzy membership function, defined as:
\begin{equation}\label{eq:Shi1}
\mu_{A_i}\left(x\right)=\frac{1}{1+\left(\frac{x-c_i}{a_i}\right)^{2b_i}},\forall i=1,2,3
\end{equation}
where ${(a}_i, b_i, c_i)$ are the parameters of the fuzzy membership function, representing the fuzzy relationship of the SNR and also referred to as antecedent parameters.

\subsubsection{Fuzzy rule layer}
This layer mainly represents the weight of each fuzzy rule. The output of this layer is:
\begin{equation}\label{eq:Shi1}
O_i^2=w_i=\mu_{A_i}x,\forall i=1,2,3
\end{equation}
where $w_i$ is the weight of the $i$-th fuzzy rule.

\subsubsection{Normalization layer}
This layer normalizes the weights of the fuzzy rules. The output of this layer can be represented as:
\begin{equation}\label{eq:Shi1}
O_i^3=\bar{w_i}=\frac{w_i}{\sum_{j=1}^{3}w_j},\forall i=1,2,3
\end{equation}

For convenience, the output of this layer is referred to as the normalized weight.

\subsubsection{Inference layer}
Each node in this layer is an adaptive node, and the output is a product of the normalized weight and a first-order polynomial (for a first-order Sugeno model). Therefore, the output of this layer is:
\begin{equation}\label{eq:Shi1}
O_i^4=y_i=\bar{w_i}\left(p_ix+q_i\right),\forall i=1,2,3
\end{equation}
where $(p_i, q_i)$ are the consequent parameters, representing the length of the prompt code under the fuzzy rule.

\subsubsection{Output layer}
\textcolor{black}{This layer is a classification node that defuzzies inference results and converts them into a probability distribution for the class. Therefore, the output of this layer is:}
\begin{equation}\label{eq:Shi1}
O_{i}^{5}=\operatorname{Softmax}\left(O_{i}^{4}\right)=\frac{e^{O_{i}^{4}}}{\sum_{j=1}^{3} e^{O_{j}^{4}}}, \forall i=1,2,3
\end{equation}

After joint encoding is completed, the encoded sequence $\mathbf{Y}$ is sent to the physical channel represented by $f:\mathbb{C}^k\rightarrow\mathbb{C}^k$. We will model the channel transmission through a series of untrained layers, represented by the transmission function $\hat{\mathbf{Y}}=f(\mathbf{Y}).$ We consider two widely used channel models: (i) Additive White Gaussian Noise (AWGN) channel, and (ii) Rayleigh fading channel. For the AWGN channel, the transmission function is $f_n(\mathbf{Y})=\mathbf{Y}+\ \mathbf{n}$, where the noise vector $\mathbf{n}\in\mathbb{C}^k$ is composed of independent and identically distributed (i.i.d.) samples from a circularly symmetric complex Gaussian distribution, i.e., $ \bm n \sim  \mathcal{CN}(0,\sigma^2\bm I_k)$, with $\sigma^2$ representing the average noise power, and $ \bm I_k$ represents the identity matrix. In the case of the Rayleigh fading channel, the channel gain's effect on the transmitted signal can be captured by the transmission function $f_h(\mathbf{Y})=\mathbf{HY}$, where $ h \sim \mathcal{CN}(0,H_c)$ is a complex random variable and \(H_c\) is the covariance matrix. The combined effect of fading and Gaussian noise can be modeled through the transmission functions $f_h$ and $f_n$: $f(\mathbf{Y})=f_n(f_h(\mathbf{Y}))=\mathbf{HY}\ +\ \mathbf{n}$ \cite{bourtsoulatze2019deep}.

The decoding process at the receiver includes joint channel-semantic-knowledge base decoding. The joint decoder will project the corrupted signal $\hat{\mathbf{Y}}=f(\mathbf{Y})\in\mathbb{C}^k$ to an estimate of the original input $\hat{\mathbf{X}}\in\mathbb{R}^n$, and knowledge base decoding performs dynamic restoration of the text based on the contextual information at the transmitter using fuzzy rules derived from the private KB.

In LGM-TSC, we combine prompt encoder/decoder, semantic encoder/decoder, and channel encoder/decoder into a joint knowledge base-semantic-channel coding scheme. The loss function of the channel codec is Mutual Information (MI) \cite{duncan1970calculation}, while the loss function of the semantic codec is CE as shown in Eq. \ref{eq:ShiCE}. Additionally, the loss function of the prompt codec is cosine similarity \cite{rahutomo2012semantic}. 
Through joint coding, our system can accurately transmit semantics, demonstrates elasticity to changes in channel conditions, and remains unaffected by sudden quality decline or ``cliff effect"\textcolor{black}{\cite{fu2024digital}}. Even when channel conditions deteriorate, our system exhibits elegant performance degradation.

Fig. \ref{fig:gkb2} illustrates the framework of joint knowledge base-semantic-channel coding, while Table \ref{tab:exam} provides an example of knowledge base coding. Initially, the prompt encoder generates a corresponding prompt for different SNRs, which is then input into the private KB. The instructions specify that, for low SNRs, the knowledge base encoding should aim to compress the text into as few characters as possible. Consequently, the private KB dynamically performs knowledge base encoding on the text according to these prompts.
For example, the original text, ``A young child, with a beaming smile, eagerly slides down the slide," would be encoded as ``A young child, smiling, slides down the slide" at low SNR levels while retaining the essential semantics. Conversely, at high SNR levels, the text remains unchanged.

\subsection{GSR for semantic reconstruction}
The GSR for semantic reconstruction consists of two key components: (1) the BERT-VITS2 model, which is responsible for reconstructing text into audio that resembles the user's timbre, and (2) the SadTalker model, which focuses on converting audio into natural, realistic talking-face video.
The above two components of GSR will work together to enable high QoE SemCom. In the following sections, we will delve into each of these components in detail.

BERT-VITS2 is a powerful Text-To-Speech (TTS) model, which accurately replicates a person's timbre in the generated audio. The workflow of audio reconstruction based on BERT-VITS2 is illustrated in Fig. \ref{fig:vits2}. For the recovered semantic text $\hat{\mathbf{T}}$ output from the private KB at the receiver and user’s vocal feature vector \textbf{v} stored in the public KB, the process of reconstructing the audio is as follows:

\begin{figure}[htbp]
	\centering
	\includegraphics[width=7.5cm]{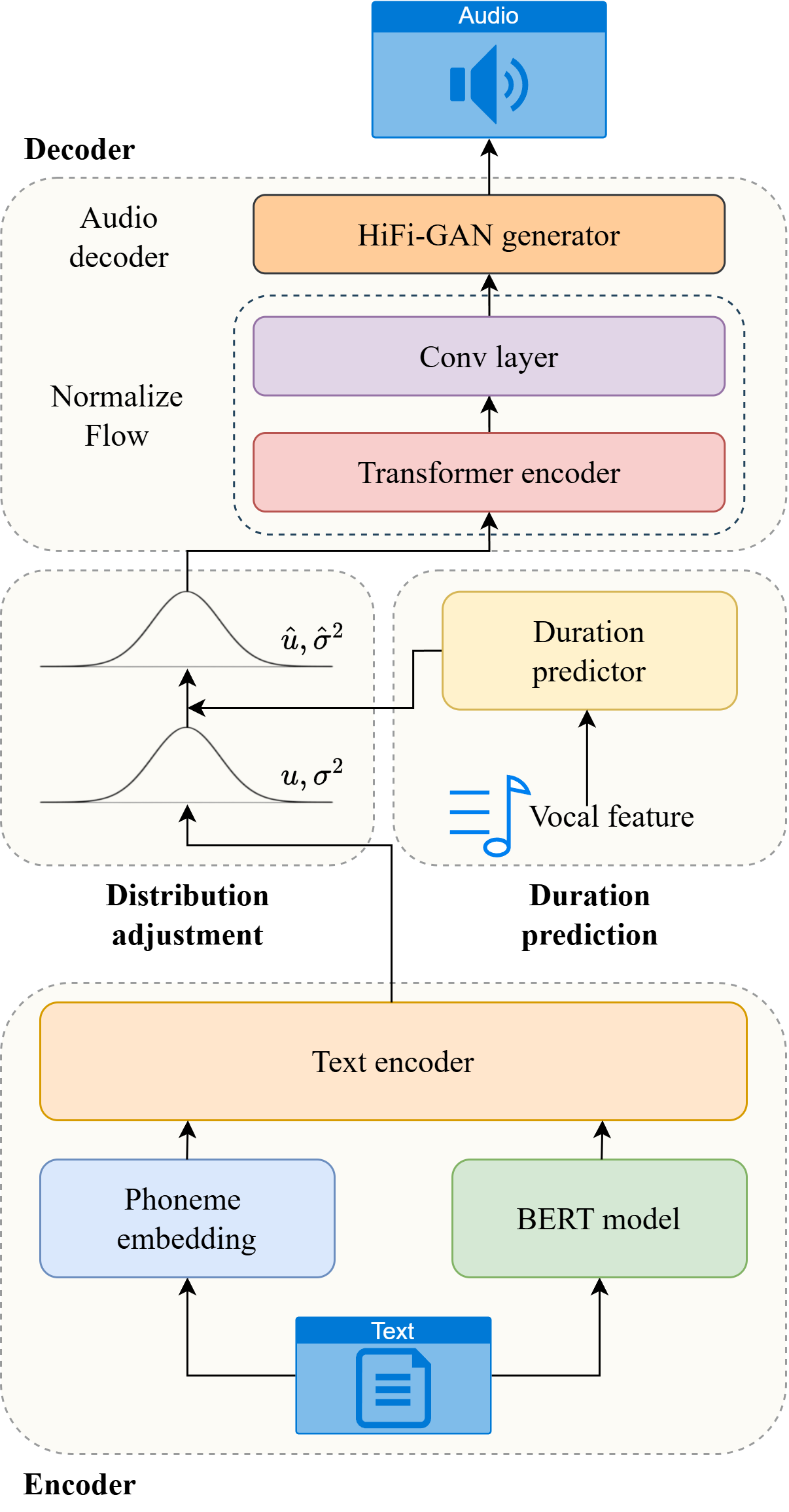}
	\caption{The architecture of the BERT-VITS2 model.}
	\label{fig:vits2}
\end{figure}

\subsubsection{Encoder}
The BERT-VITS2 model not only encodes phonemes but also uses a BERT model to pre-encode the text. This encoded text is then combined with the phoneme sequence as input to the text encoder based on the transformer encoder, and outputting the hidden state $\mathbf{T}_{hid}$ of the text as follows:
\begin{equation}\label{eq:18}
\mathbf{T}_{hid}=\mathrm{ENC}\left(\mathrm{B}\left(\hat{\mathbf{T}}\right),\mathrm{P}\left ( \hat{\mathbf{T}} \right )  \right)
\end{equation}
where \(\mathrm{P}\left( \cdot\right) \) is the phoneme embedding matrix, which is used to capture details such as rhythm and emotion contained in \(\hat{\mathbf{T}}\). $\mathrm{B}\left(\cdot\right)$ is the BERT model, \textcolor{black}{and $\mathrm{ENC}\left(\cdot\right)$ is the text encoder which is used to synthetically encode the text and its phoneme.}

\subsubsection{Duration prediction}
The BERT-VITS2 model uses a Duration predictor based on convolutional layers and a random forest \cite{breiman2001random} to predict the duration of each phoneme, serving as a reference for generating audio. This module uses the hidden state of the text $\mathbf{T}_{hid}$, user's vocal feature vector $\mathbf{v}$, and a random noise \textcolor{black}{$\textbf{N}$} to output the duration $\mathbf{T}_{dp}$ of each phoneme:
\begin{equation}\label{eq:19}
\mathbf{T}_{dp}=\mathrm{DP}\left(\mathbf{T}_{hid},\mathbf{v},\mathbf{N}\right)
\end{equation}
where $\mathrm{DP}\left(\cdot\right)$ is the Duration predictor.

\subsubsection{Distribution adjustment}
The BERT-VITS2 model maps $\mathbf{T}_{hid}$ to a distribution with a certain mean \(\mu \) and variance \(\sigma ^2\). We aim to derive the distribution from the text feature to approximate the distribution of real audio features, i.e., the distribution of real acoustic features. Therefore, the BERT-VITS2 model dynamically adjusts this distribution to the \textcolor{black}{mean \(\hat{\mu}\) and variance \(\hat{\sigma} ^2\)} based on the output duration $\mathbf{T}_{dp}$ from the Duration predictor to produce speech with natural prosody.

\subsubsection{Decoder}
A set of features can be randomly sampled from the above distribution to serve as the subsequent input, denoted as $\mathbf{z}_s$. Before decoding, these features need to pass through a Normalizing Flow module. Then the audio decoder projects these features into the original waveform of the audio, ultimately synthesizing the audio:
\begin{equation}\label{eq:20}
\mathbf{A}=\mathrm{DEC}(\mathrm{FLOW}(\mathbf{z}_s))
\end{equation}
\textcolor{black}{where $\mathrm{FLOW}\left(\cdot\right)$ is the Normalizing Flow module which includes the transformer encoder and convolutional layer to enhance the model's expressive power \cite{prenger2019waveglow}, and $\mathrm{DEC}\left(\cdot\right)$ is the HiFi-GAN generator audio decoder\cite{kong2020hifi} which can capture long-range dependencies and complex acoustic characteristics. \textcolor{black}{$\mathbf{A}$} is the final output audio synthesized by the model.}

SadTalker is a generative model designed to produce realistic speaking face videos from audio and a single-face image. 
It consists of several key components, including ExpNet, PoseVAE, and Face Render. 
SadTalker utilizes ExpNet and PoseVAE to get 3D model expression coefficients and head postures, respectively, resulting in videos that are state-of-the-art in terms of lip sync and video quality. 
At the receiver, we employ the SadTalker model as the component to reconstruct the talking-face video in GSR. The workflow of video reconstruction based on SadTalker is shown in Fig. \ref{fig:sd}. For the user's face image $f$ stored in the public KB and the $\mathbf{A}$ reconstructed by BERT-VITS2, the process of rebuilding the video is as follows:

\begin{figure}[htbp]
	\centering
	\includegraphics[width=8.5cm]{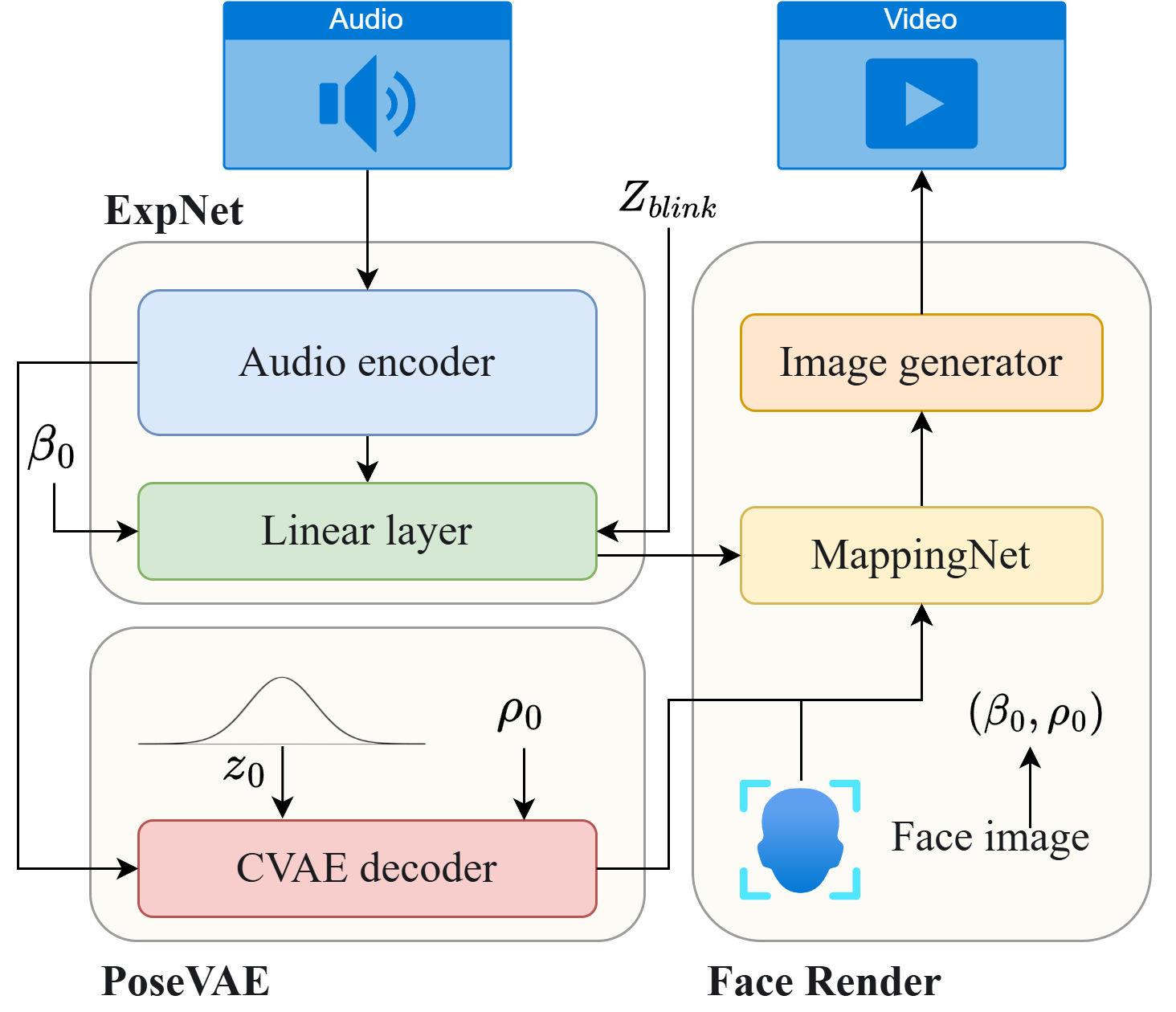}
	\caption{The architecture of the SadTalker model.}
	\label{fig:sd}
\end{figure}

\setcounter{subsubsection}{0}
\subsubsection{ExpNet}
\textcolor{black}{
	The SadTalker model first analyzes the input image \textcolor{black}{\(\textbf{P}\)} to obtain the initial emotion coefficient \(\beta _0\) and head pose motion coefficient \(\rho _0\). 
	These coefficients are then used to predict the future frames. 
	The ExpNet module uses a ResNet-based\cite{he2016deep} audio encoder $\mathrm{\Phi}_A$ to encode \(\mathbf{A}\) into an audio embedding. This audio embedding is then input to a linear layer $\mathrm{\Phi}_M$, outputting the emotion coefficient set $\beta_{exp}$ of the video:}
\begin{equation}\label{eq:21}
\beta_{exp}={\mathrm{\Phi}_M(\mathrm{\Phi}}_A\left(\mathbf{A}\right),z_{blink},\beta_0)
\end{equation}
where $z_{blink}\in[0,1]$ is a randomly generated blink control signal, and $\beta_0$ is the initial emotion coefficient extracted from \textcolor{black}{\(\textbf{P}\)}.

\subsubsection{PoseVAE}
The SadTalker model uses the PoseVAE module to generate head pose motion coefficients. 
	The PoseVAE module is based on a Conditional Variational AutoEncoder (CVAE) decoder \cite{sohn2015learning}. These coefficients describe the rotation and tilt of the head. First, a Gaussian distribution sample $z_0$ is taken, and the first frame head pose motion coefficient and audio features from ExpNet serve as conditions. The decoder outputs the head pose motion coefficient set $\rho$ of the video, formulated as follows:
\begin{equation}
	\rho =\mathrm{CVAE} \left ( \Phi _A\left ( \mathbf{A} \right )  \right ) 
\end{equation}
\textcolor{black}{where \(\mathrm{CVAE}\left( \cdot \right) \) is the CVAE decoder, and \(\rho\) follows the following distribution:}
\begin{equation}\label{eq:22}
p_\phi\left(\rho|z_0,c\right)=N\left(\rho\middle|\mu_\phi\left(z_0,c\right),\sigma_\phi\left(z_0,c\right)^2\mathbf{I}\right)
\end{equation}
where $\phi$ is the parameter set of the decoder, $\mu_\phi$ and $\sigma_\phi$ are the modules inside CVAE that generate the mean and variance respectively, $\mathbf{I}$ is the identity matrix, and $c$ is the condition for the decoding process, including the audio features and the first frame head pose motion coefficient $\rho_0$.

\subsubsection{Face Render}
Finally, SadTalker utilizes the Face Render module to compose the video. The facial expression coefficient set $\beta_{exp}$ and head motion coefficient set $\rho$ generated by the ExpNet and PoseVAE modules are combined to form a complete 3D motion coefficient set. The Face Render module employs the one-dimensional convolution-based MappingNet module to map this 3D motion coefficient to the 3D keypoint space. Subsequently, the Face Render module utilizes the image generator based on 3D convolution, gradually generating each frame of the video stream image based on the original face image and the mapped coefficients. Ultimately, in chronological order, the model combines these generated video frames to form a continuous video stream.

The workflow of GSR is illustrated in \textbf{Algorithm} \ref{alg2}.

\begin{algorithm}
	\caption{GSR}
	\label{alg2}
	\begin{algorithmic}[1]
		\REQUIRE Recovered semantic text $\hat{\mathbf{T}}$, user's face image $\textbf{P}$ and vocal feature vector $\mathbf{v}$
		\ENSURE Reconstructed talking-face video $\hat{\mathbf{X}}$
		\STATE{Convert $\hat{\mathbf{T}}$ back into corresponding audio $\mathbf{A}$ according to Eqs. (\ref{eq:18})-(\ref{eq:20})}.
		\STATE{Reconstruct the $\hat{\mathbf{X}}$ using the $\mathbf{A}$, \textcolor{black}{\(\textbf{P}\)} and \textbf{v} according to Eqs. (\ref{eq:21})-(\ref{eq:22}).}
		\STATE{Return $\hat{\mathbf{X}}$.}
	\end{algorithmic}
\end{algorithm}

\section{Numerical Results}
In this section, we evaluate the performance of the proposed LGM-TSC system by comparing it with other SemCom systems and some traditional video codec algorithms.

\subsection{Simulation settings}

The experimental setup is as follows:
The semantic encoder is a pre-trained BERT model with 768 feature dimensions. The channel encoder is a stacked autoencoder with an input dimension of 768 and an output dimension of 256. The channel decoder is also a stacked autoencoder, with an input dimension of 256 and an output dimension of 768. The semantic decoder is the final linear layer of the BERT model, which has an input dimension of 768 and an output dimension corresponding to the vocabulary size.

The experimental training and testing environments include Windows 11, Python 3.8, PyTorch 2.0.1, and CUDA 11.8. Computing resources consist of a 12th Gen Intel® Core™ i7-12700H processor running at 2.30 GHz, along with NVIDIA GeForce RTX 4070 GPUs.

%\subsection{Evaluation metrics}

%To evaluate the performance of the LGM-TSC system, we design three corresponding indicators:
%(1) text level: assessing the compression ratio of the GSE in converting video to text;
%(2) semantics level: evaluating whether the private KB impacts the downstream task performance of the received text data under varying SNRs;
%(3) video level: following the approach outlined in Txt2Vid \cite{tandon2022txt2vid}, we evaluated the quality of the reconstructed video through subjective assessments by 50 participants. 

\subsection{Performance evaluation for GSE}
\begin{figure}[htbp]
	\centering
	\includegraphics[width=8.5cm]{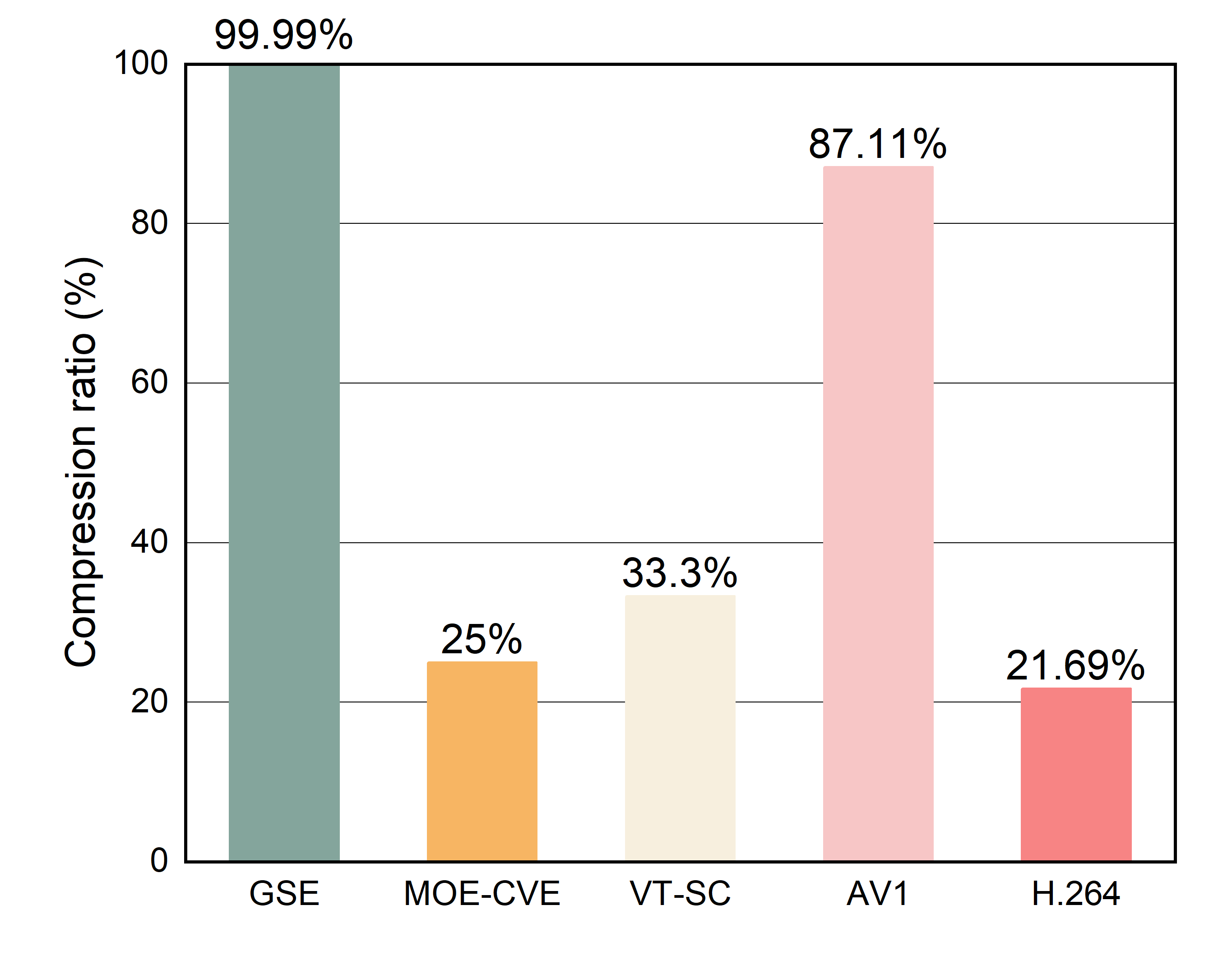}
	\caption{\textcolor{black}{Compression ratio comparison of the GSE with other algorithms.}}
	\label{fig:cse}
\end{figure}

To assess the performance of GSE, we conducted a comparison with two traditional video compression algorithms, H.264\cite{wiegand2003overview} and AV1\cite{chen2018overview}, as well as two other video SemCom systems, MOE-CVE\cite{li2024video} and VT-SemCom\cite{zhang2023deep}, in terms of compression ratio.
The compression ratio is defined as the proportion of data that has been compressed, calculated
by dividing the amount of data reduced during
compression by the size of the original data. This
means that less transmitted data indicates a higher
compression rate.
The experimental results are presented in Fig. \ref{fig:cse}.

As depicted in Fig. \ref{fig:cse}, the LGM-TSC system achieves the highest compression ratio for GSE. AV1 follows closely behind, while H.264 and the other two SemCom systems exhibit lower rates. These findings indicate that the GSE module effectively converts bandwidth-intensive talking-face videos into text, reducing the amount of data transmitted by four orders of magnitude compared to direct transmission of raw video. This performance surpasses that of other direct coding methods based on deep learning and traditional codec approaches.

%Our results show that GSE has a compression advantage of up to 100 times over transmitting raw video directly.

\subsection{Performance evaluation for the private KB}
\begin{figure}[htbp]
	\centering
	\includegraphics[width=8.5cm]{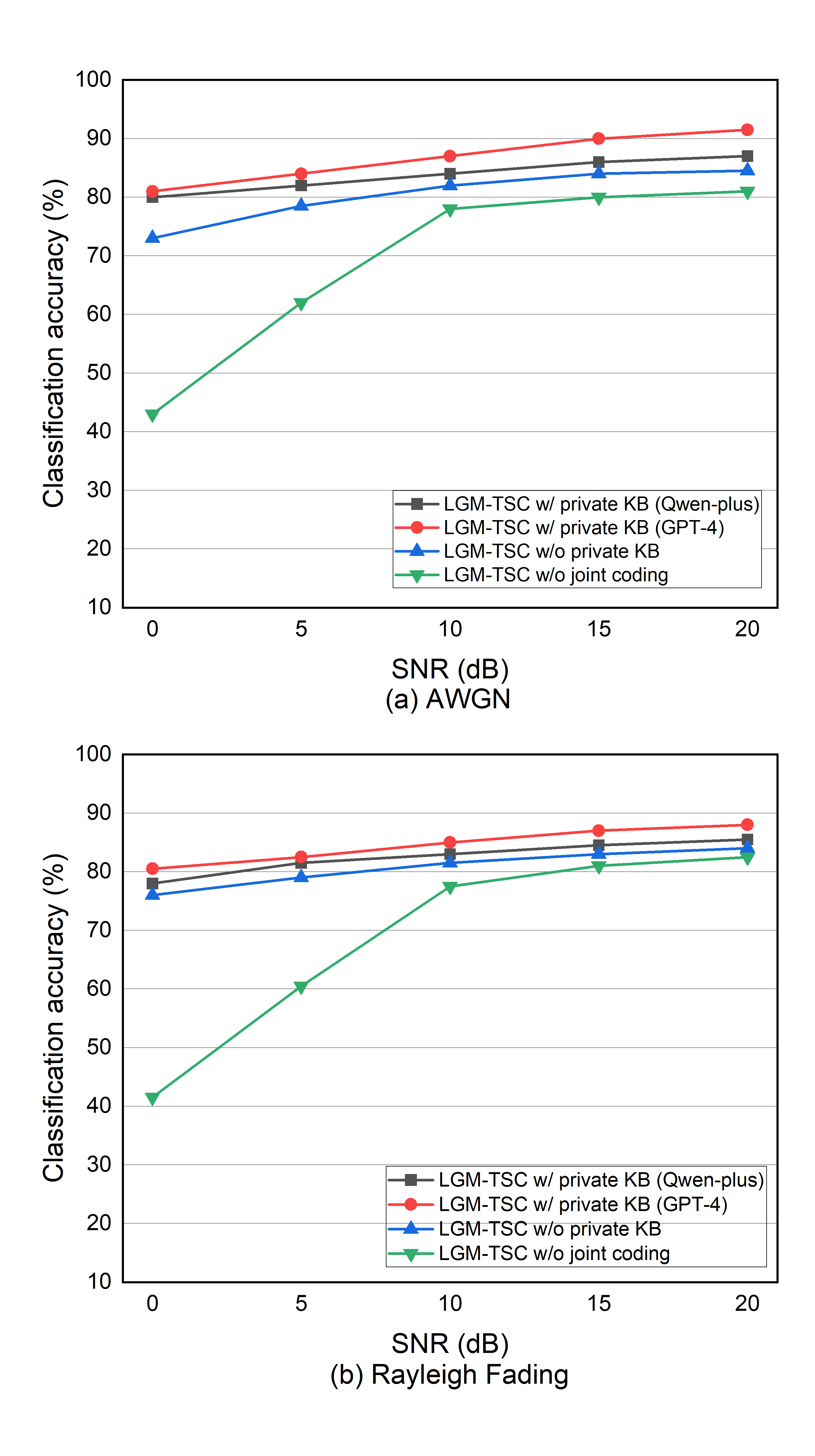}
	\caption{
Ablation study of the LGM-TSC system and its performance in downstream classification tasks}
	\label{fig:gkb}
\end{figure}

In order to evaluate the performance of the private KB in the SemCoM system, we respectively apply the receiver text without the private KB and with the private KB to the same text classification task in two channel environments (AWGN and Rayleigh), \textcolor{black}{and compared the performance of downstream classification tasks using the same pre-trained model on these texts.} The metric used for performance evaluation is classification accuracy. The selected dataset is the IMDB dataset\cite{maas2011learning}. The LLM for the private KB application is GPT-4 and Qwen-plus \cite{qwen} respectively. The experimental results are shown in Fig. \ref{fig:gkb}.

As depicted in Fig. \ref{fig:gkb}, \textcolor{black}{we compared LGM-TSC with Qwen-plus and GPT-4 as private KB, LGM-TSC without private KB, and LGM-TSC without joint coding, respectively.} It is evident that the presence of the private KB significantly enhances the performance of transmitted data in downstream tasks at the receiver across various channel conditions, especially when the SNR is low. Even under high SNR conditions, the SemCom system with the private KB slightly outperforms the system without it, highlighting the private KB's crucial role in mitigating transmission noise. This can be attributed to the LLM's robust natural language understanding, which dynamically adjusts text based on SNR to ensure the accurate transmission of the original semantics. Additionally, a comparison with a SemCom system without joint coding shows that the LGM-TSC system effectively overcomes the ``cliff effect" after joint knowledge base-semantic-channel coding under random SNR conditions.
\vspace*{-3mm} 
\subsection{Performance evaluation for GSR}
\begin{figure*}[htbp]
	\centering
	\includegraphics[width=17cm]{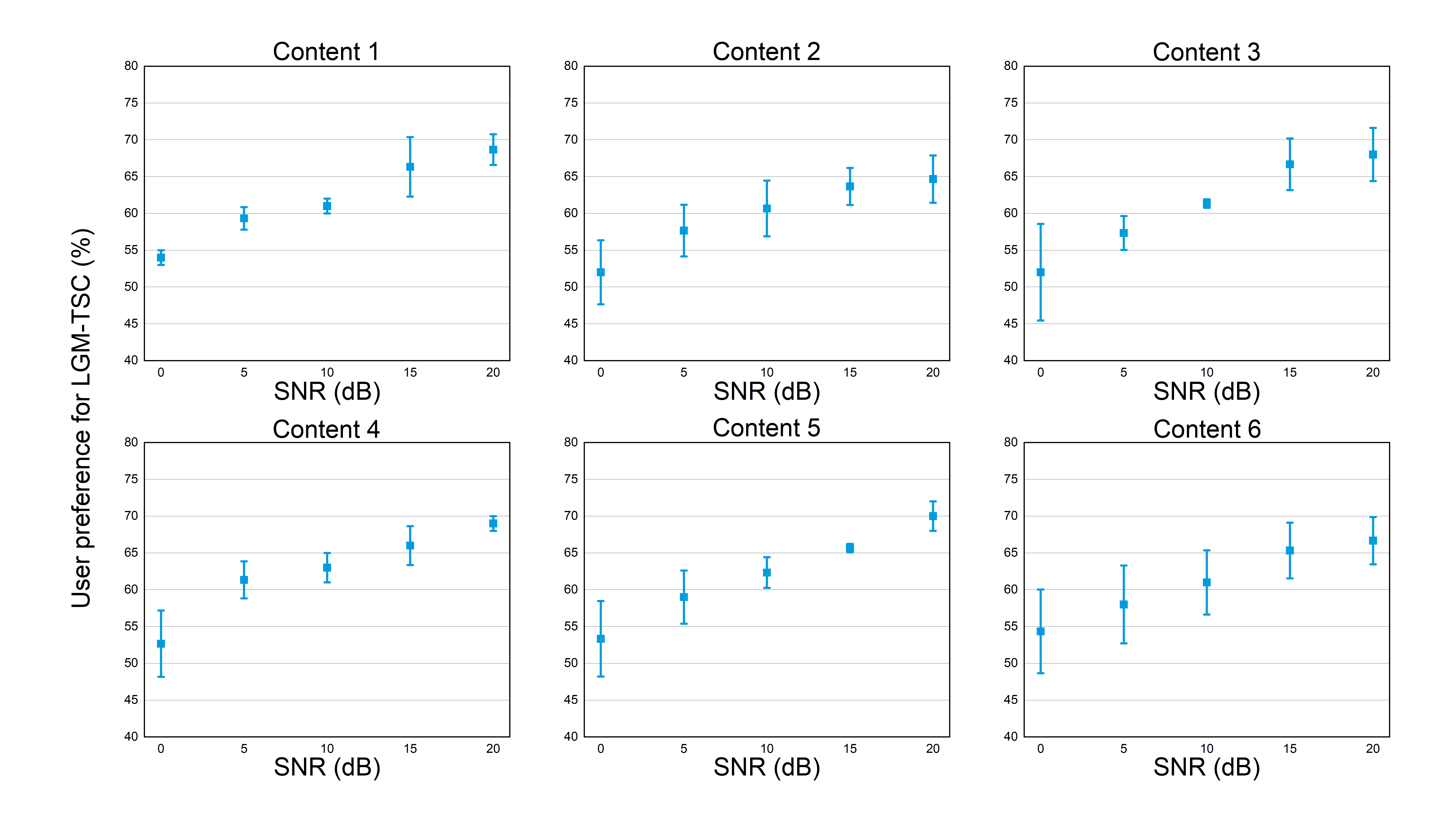}
	\caption{\textcolor{black}{Subjective study results: comparing videos reconstructed by LGM-TSC system to standard codecs (including H.264 and AV1). Each plot shows the percentage of users who prefer the LGM-TSC over H.264 and AV1 under different SNRs. The error bars show 95\% confidence interval.}}
	\label{fig:csr}
\end{figure*}

%To assess the QoE of the video reconstructed by GSR, we create a new dataset by combining the videos reconstructed by GSR at different SNRs with the video codecs H.264 and AV1. Subsequently, 50 participants were asked to select the video with the best QoE based on their subjective perceptions. We then calculate the proportion of participants who chose the LGM-TSC system among the four groups of videos in the dataset. The experimental results are presented in Fig. \ref{fig:csr}. It should be noted that errors may have occurred due to participants' inability to accurately identify optimal QoE.
The dataset employed in this experiment to assess the QoE of videos reconstructed is a facial video dataset provided for the Txt2Vid Subjective Study \cite{tandon2022txt2vid}. It includes footage of six different individuals (spanning various ethnicities; four men and two women) under a range of natural indoor lighting and speaking conditions. We calculate the proportion of participants who choose the LGM-TSC system among the six groups of videos in the dataset. The experimental results are presented in Fig. \ref{fig:csr}. %It is important to note that errors may have occurred due to participants' potential difficulty in accurately identifying optimal QoE. 

\begin{figure}[htbp]
	\centering
	\includegraphics[width=8.5cm]{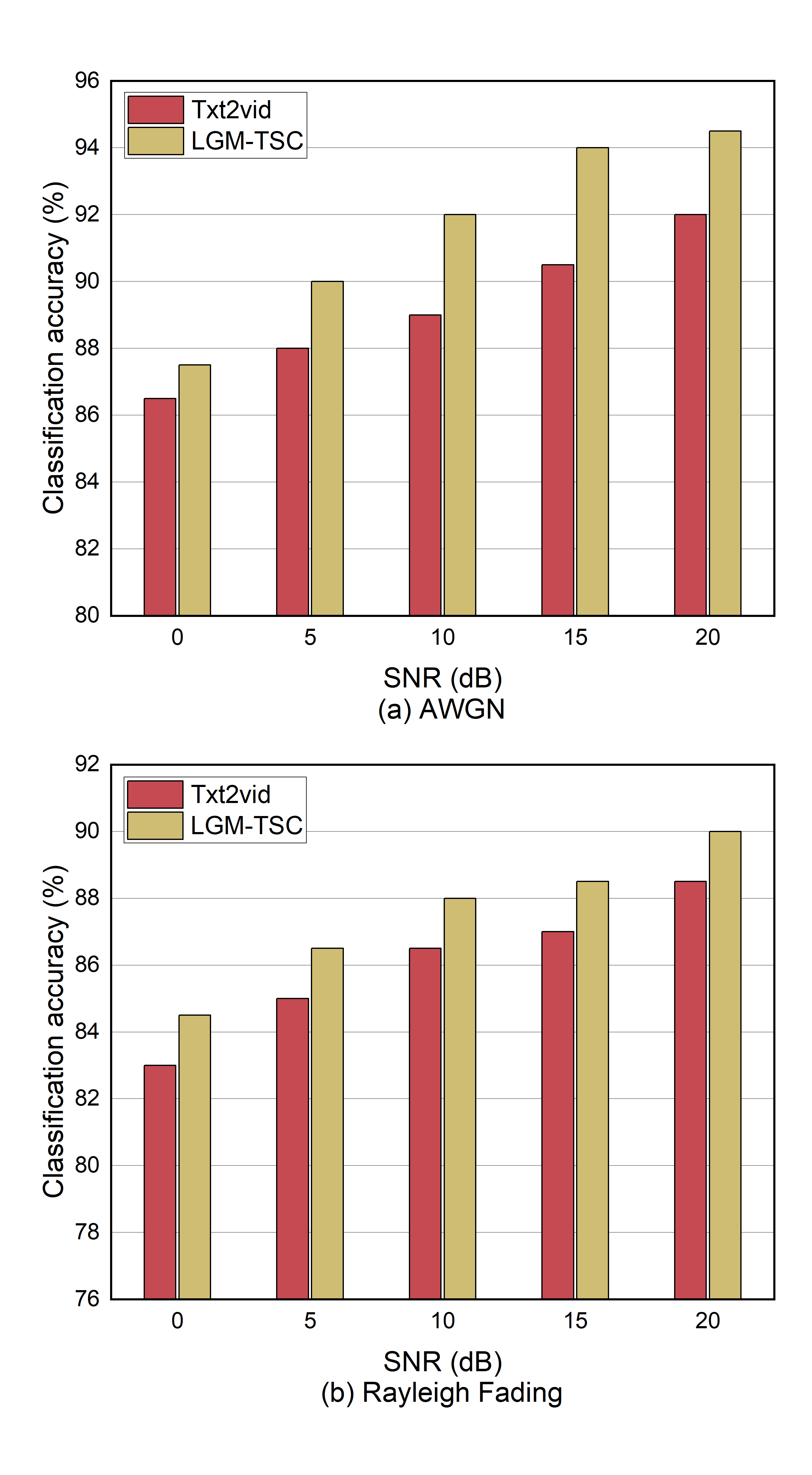}
	\caption{Performance comparison of LGM-TSC and Txt2vid in downstream classification  tasks.}
	\label{fig:sc}
\end{figure}

Fig. \ref{fig:csr} shows that more than half of the participants preferred LGM-TSC reconstructed videos across different SNRs, with the selection ratio increasing as the SNR improves, indicating that videos reconstructed by GSR provide a sufficiently high QoE. This can be attributed to the precise timbre cloning achieved by the BERT-VITS2 models and the robust face-talking modeling technology of the SadTalker models, with higher SNR levels further enhancing users' comprehension of the semantics.

\subsection{SemCom performance evaluation}
\begin{figure*}[htbp]
	\centering
	\includegraphics[width=17cm]{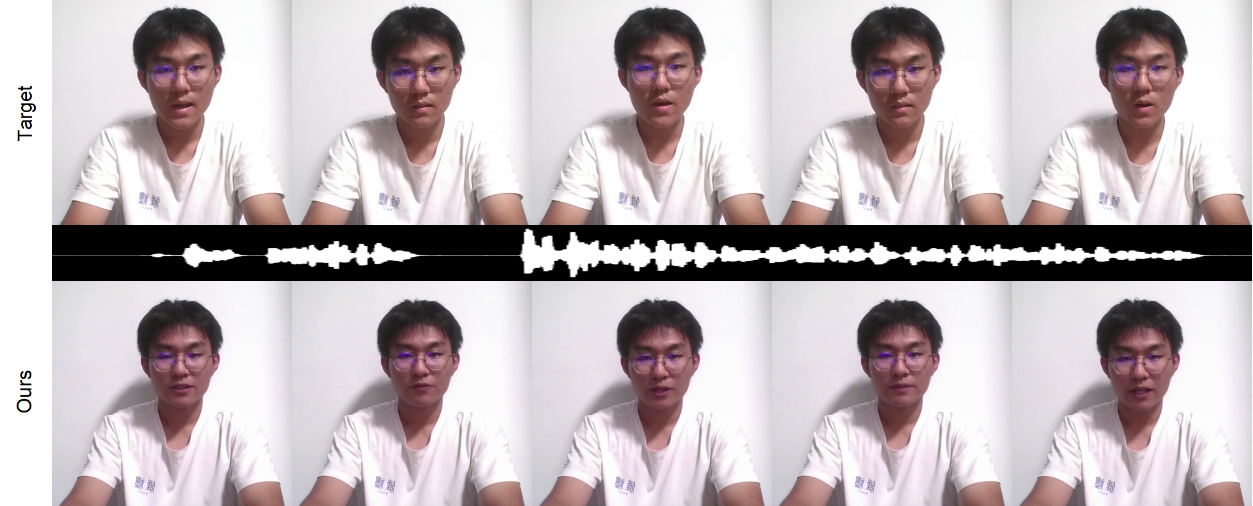}
	\caption{Multiple frames belonging to the original video (Target) and the reconstructed video (Ours) by the LGM-TSC, which exhibit a similar QoE to the original in terms of lip-sync accuracy, video dynamics, and timbre similarity.}
	\label{fig:compare}
\end{figure*}

\textcolor{black}{To evaluate the performance of the LGM-TSC system in the downstream classification task, we compare it with the Txt2vid system \cite{tandon2022txt2vid}. The metric used for performance evaluation is classification accuracy. The selected dataset is the Sms\_spam dataset \cite{almeida2011contributions}, and the pre-trained model for the downstream task is the BERT-base model.} 
The LLM used in the private KB is GPT-4.
The experimental results are shown in Fig. \ref{fig:sc} and Fig. \ref{fig:compare}.

Fig. \ref{fig:sc} demonstrates that, across various channel environments, LGM-TSC slightly outperforms the Txt2vid system in classification tasks at low SNR, while both systems show similar performance at high SNR. This advantage arises because a joint coding incorporating the private KB can effectively resist channel noise and transmit more accurate semantics.
Fig. \ref{fig:compare} displays multiple frames from both the original video and the LGM-TSC reconstruction result. It is evident that the reconstructed video maintains picture quality comparable to the original video, with lip-sync accuracy nearly matching that of the original, aligning well with the video dynamics and timbre similarity.
Thus, we conclude that LGM-TSC can transmit talking-face video both accurately and efficiently. This effectiveness can be attributed to the following factors: first, GSR's robust generation capability ensures realistic and dynamic video reconstruction; second, the private KB and joint coding guarantees accurate semantic transmission, resulting in the synchronization of the reconstructed video with the original video.

\section{Conclusions}
In this paper, we introduce a novel LGM-TSC system to compress talking-face video into text for transmission over wireless channels and subsequently reconstruct the video at the receiver. The system consists of three primary components: the GSE for compressing video into text, the private KB for text disambiguation and correction, and the GSR for high-fidelity talking-face video reconstruction. Additionally, we propose a joint knowledge base-semantic-channel coding method that enhances the performance of the SemCom system. Experiments conducted on portrait video datasets demonstrate that the LGM-TSC system achieves ultra-low bitrate compression without compromising video semantics.

\bibliographystyle{IEEEtran}
\bibliography{bare_jrnl_bobo}

\newpage
\end{document}